\documentclass[prb,preprint]{revtex4}

\usepackage{graphicx}
\usepackage[T1]{fontenc}
\usepackage[latin9]{inputenc}
\usepackage{verbatim}
\usepackage{amsmath}
\begin{document}

\title{Molybdenum sputtering film characterization for high gradient accelerating structures}
\author{S. Bini, B. Spataro}
\affiliation{INFN - LNF, Via E. Fermi 40, 00044 Frascati (RM), Italy}
\author{A. Marcelli}
\affiliation{INFN - LNF, Via E. Fermi 40, 00044 Frascati (RM), Italy, NSRL, University of Science and Technology of China, Chinese Academy of Sciences, Hefei, Anhui 230029, P.R. China}
\author{S. Sarti}
\affiliation{University of Rome - Sapienza, Dipartimento di Fisica, Piazzale Aldo Moro 5, 00185 Rome, Italy}
\author{V.A. Dolgashev, S. Tantawi, A.D. Yeremian}
\affiliation{LAC National Accelerator Laboratory, 2575 Sand Hill Road, Menlo Park, CA 94025, USA}
\author{Y. Higashi}
\affiliation{KEK 1-1 Oho, Tsukuba, Ibaraki 305, Japan}
\author{M.G. Grimaldi, L. Romano, F. Ruffino}
\affiliation{University of Catania, Dipartimento di Fisica e Astronomia \& MATIS-IMM-CNR, Via S. Sofia 64, 95123 Catania, Italy}
\author{R. Parodi}
\affiliation{INFN-Genova, Via Dodecaneso 33, 16146 Genova, Italy}
\author{G. Cibin}
\affiliation{Diamond Light Source, Chilton, Didcot, Oxford OX110DE, UK}
\author{C. Marrelli, M. Migliorati}
\affiliation{University of Roma - Sapienza, Dipartimento di Scienze di Base e Applicate per l'Ingegneria, Via A. Scarpa, 14 - 00161 Roma - Italy}
\author{C. Caliendo}
\affiliation{Istituto dei Sistemi Complessi, ISC-CNR, Are della Ricerca di Roma 1, Via Salaria km 29.300, 00040 Monterotondo, Rome, Italy}

\begin{abstract}
Technological advancements are strongly required to fulfill the demands of new accelerator devices with the highest accelerating gradients and operation reliability for the future colliders. To this purpose an extensive R$\&$D regarding molybdenum coatings on copper is in progress. In this contribution we describe chemical composition, deposition quality and resistivity properties of different molybdenum coatings obtained via sputtering. The deposited films are thick metallic disorder layers with different resistivity values above and below the molibdenum dioxide reference value. Chemical and electrical properties of these sputtered coatings have been characterized by Rutherford backscattering, XANES and photoemission spectroscopy. We will also present a three cells standing wave section coated by a molybdenum layer $\sim$ 500 nm thick designed to improve the performance of X-Band accelerating systems. 
\end{abstract}

\maketitle

\newcommand{\rmi}{\mathrm{i}}
\newcommand{\etal}{{\it et al}}

\newcommand{\IJMPB}{Int. J. Mod. Phys. B }
\newcommand{\PC}{Physica C }
\newcommand{\PB}{Physica B }
\newcommand{\JS}{J. Supercond. }
\newcommand{\IEEEmw}{IEEE Trans. Microwave Theory Tech. }
\newcommand{\IEEEas}{IEEE Trans. Appl. Supercond. }
\newcommand{\IEEEim}{IEEE Trans. Instr. Meas. }
\newcommand{\PRB}{Phys. Rev. B }
\newcommand{\PRL}{Phys. Rev. Lett. }
\newcommand{\PR}{Phys. Rev. }
\newcommand{\PL}{Phys. Lett. }
\newcommand{\IJIMW}{Int. J. Infrared Millim. Waves }
\newcommand{\APL}{Appl. Phys. Lett. }
\newcommand{\JAP}{J. Appl. Phys. }
\newcommand{\JPCM}{J. Phys.: Condens. Matter }
\newcommand{\JPCS}{J. Phys. Chem. Solids }
\newcommand{\AdP}{Adv. Phys. }
\newcommand{\Nat}{Nature }
\newcommand{\CM}{cond-mat/ }
\newcommand{\JpnJAP}{Jpn. J. Appl. Phys. }
\newcommand{\PhT}{Phys. Today }
\newcommand{\ZETF}{Zh. Eksperim. i. Teor. Fiz. }
\newcommand{\JETP}{Soviet Phys.--JETP }
\newcommand{\EL}{Europhys. Lett. }
\newcommand{\Sci}{Science }
\newcommand{\EJPB}{Eur. J. Phys. B }
\newcommand{\IJMB}{Int. J. of Mod. Phys. B }
\newcommand{\RPP}{Rep. Prog. Phys. }
\newcommand{\SUST}{Supercond. Sci. Technol. }
\newcommand{\JLTP}{J. Low Temp. Phys. }
\newcommand{\RSI}{Rev. Sci. Instrum. }
\newcommand{\RMP}{Rev. Mod. Phys. }
\newcommand{\LTP}{Low Temp. Phys. }
\newcommand{\CPC}{Chin. Phys. C }

\section{Introduction}

The next generation of linear accelerators is highly demanding in terms of accelerating gradients. Indeed, the accelerating gradient is one of the major parameters of a linear accelerator since it determines the accelerator length and its power consumption. Although the accelerating gradient of normal conducting accelerating structures is mainly limited by the RF breakdown, the experience shows that obtaining high gradients with a normal-conducting structure requires operation at a relatively high frequency. To minimize costs, the availability of advanced high frequency accelerating structures is one of the main goals of the accelerators community. 
To upgrade performances of the X band Linacs at 11.424 GHz, resources and efforts are then devoted to achieve the highest accelerating gradients and at the same time the highest possible reliability. In the framework of a large collaboration among SLAC (Stanford Linear Accelerator Center), KEK (K$\bar{o}$ Enerug$\bar{i}$ Kasokuki Kenky$\bar{u}$ Kik$\bar{o}$) and the INFN-LNF laboratory, efforts are devoted to design, manufacture and test short high power standing wave (SW) sections.\\

Among the many, manufacture technologies and surface properties are fundamental issues of advanced cavities. Electroplating is a very attractive technique to manufacture compact structures avoiding the soft brazing while maintaining mechanical properties and high-vacuum requirements. Niobium film technology is also a well-established technology providing considerable benefits over bulk niobium. The reliable performance of niobium thin film cavities allows their use at LEP \cite{IPAC10} and LHC \cite{PRST09} although it is not clear if this technology should be used for new accelerators such as ILC (\cite{IPAC10} and therein). Molybdenum is an interesting material for accelerator components and a stimulating option for RF linear accelerating structure characterized by low breakdowns at high RF power. For application in high gradient accelerating structures, data of Mo breakdown rate and its electric field are already very promising if compared with Cu materials \cite{bini11}. 
A brazed molybdenum bulk X-band RF structure has also recently manufactured and tested at high power at SLAC. This brazed Mo structure with the same RF parameters exhibited a breakdown rate higher than a similar copper structure. However, a drawback of the manufacturing of sintered molybdenum bulk is the long time required to machining the cavity, a 300 nm surface roughness using "tungsten carbide" tools, the gas contamination and an uneven loading stress in the braze region. Additionaly technological problems have been also detected in sintered molybdenum bulk brazed sections \cite{spataro11, Higashi}. To overcome drawbacks associated to the use of sintered molybdenum bulk \cite{Ressler02}, we started a feasibility study with a single sputtering technique to make copper cavity resonators characterized by a low surface roughness and then to deposit a molybdenum film on top in order to reduce the breakdown phenomenon. A molybdenum sputtering method has been then tested as a possible route to achieve accelerating electric field performance of normal conducting structures operating at 11.424 GHz. This research will be certainly useful to improve properties of normal conducting devices and the related manufacture technologies. Although, as it will be described later, setting of Mo deposition parameters is not the main issue, a lot of work is necessary to fully characterize these devices and, in particular, their conductivity properties and behavior under high fields because because future dedicated RF cavities have to be manufactured and tested at higher power. \cite{BiniCPC} 
We present here a characterization of high-quality Mo coatings on Cu, their chemical composition and resistivity measurements with the final goal to design and manufacture a SW structure made of copper cells coated with Mo.

\section{Molibdenum coatings on copper}

We performed this study at the Laboratori Nazionali di Frascati of the INFN where we recently set up a facilities for surface coating treatments. We used copper samples of 2 cm of diameter, 3 mm thick on top of which we deposited a molybdenum film.  The method is based on a magnetron sputtering configuration. Studies shown that molybdenum thin films are easily deposited by RF magnetron sputtering (a less sensitive technique), achieve a good adhesion to the substrate and, with the same conditions of argon pressure and RF power, are characterized by a better surface quality than similar coating deposited by the DC magnetron sputtering method. Also the grain sizes of the deposited films grow with similar parameters.
A DC glow discharge procedure has been adopted before each treatment in order to clean the inner surfaces of the vacuum chamber and remove water from the environment. The glow discharge was stabilized at 1.5 A resulting in a potential of $\sim$200 V between the central molybdenum anode and the grounded vacuum chamber. The RF sputtering discharge was established in a noble gas atmosphere (argon for standard coatings) in the pressure range of $1\div15 \times 10^{-3}$ mbar. The coating took place at room temperature and a thickness of about 0.7 $\mu$m was typically obtained after 25 minutes. The distance between samples and the molybdenum target is about 8 cm at the magnetic induction of about 100 G.
 
\begin{figure}[h]
\centerline{\includegraphics[scale=0.8]{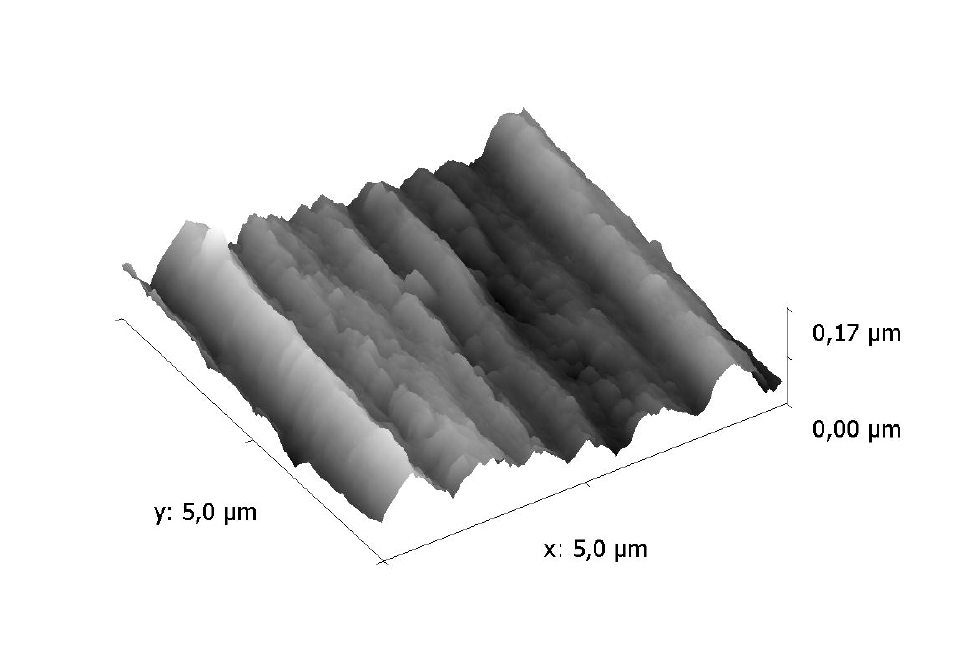}}
\caption{AFM image of the surface of a copper sample before the molybdenum sputtering.}
\label{fig:Fig1}
\end{figure}

\begin{figure}[h]
\centerline{\includegraphics[scale=0.8]{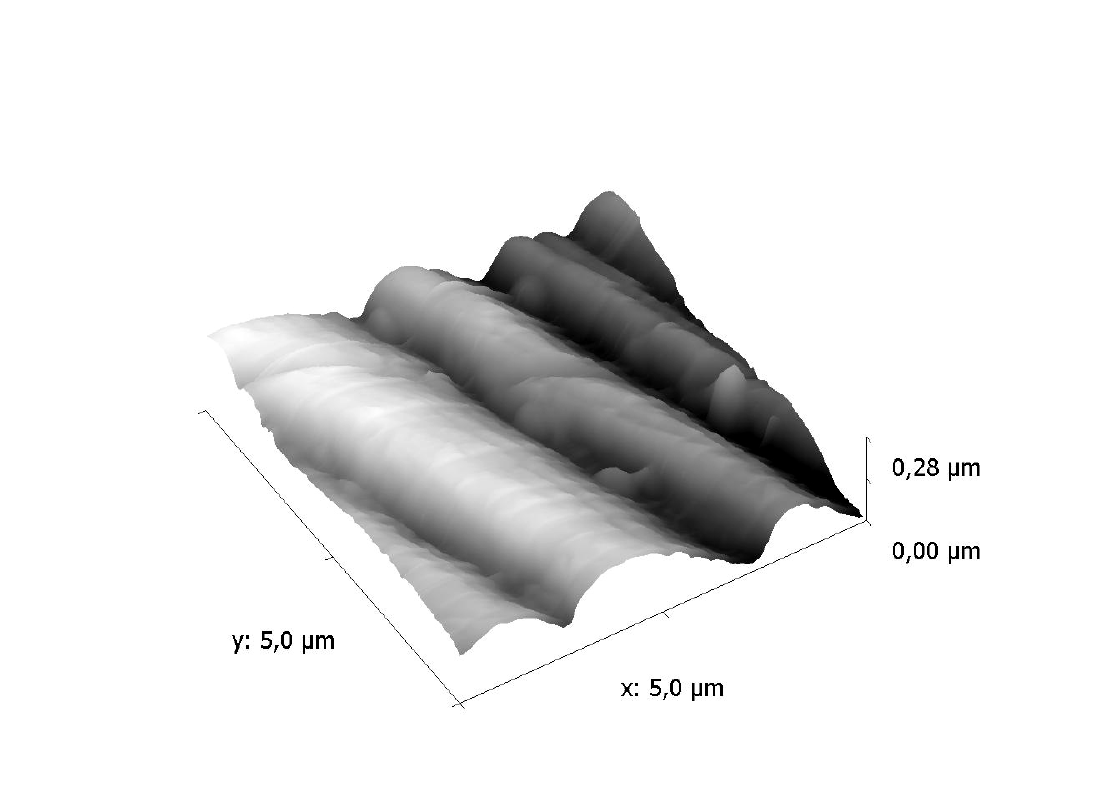}}
\caption{AFM image of deposited molybdenum on copper by the sputtering technique.}
\label{fig:Fig2}
\end{figure}

\begin{figure}[h]
\centerline{\includegraphics[scale=0.8]{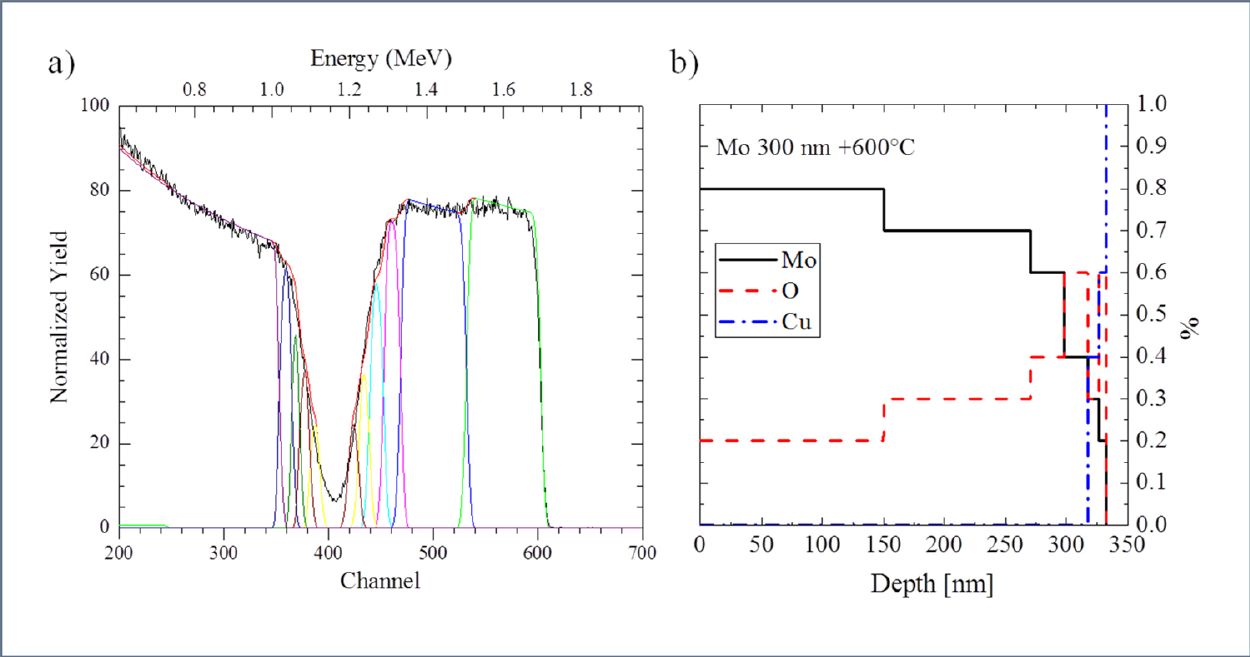}}
\caption{a) 2 MeV He RBS spectrum of Mo film deposited on Cu (black line). The RUMP
simulation of the whole sample (red line) has been obtained with different layers (colored lines) containing Mo, O and Cu. b) The resulting profiles of Mo, O and Cu as a function of depth.}
\label{fig:Fig3}
\end{figure}

To evaluate the surface quality and to understand the morphological aspects we used an atomic force microscopy (AFM). In figure \ref{fig:Fig1} we show one surfaces of the machined copper with a roughness of $\sim$70 nm before the molybdenum sputtering process. Clear undulations of the surface due to the step of lathe machine ($\sim$700 nm) with many overlapping spikes appear. 
In figure \ref{fig:Fig2} the same spikes disappear after a deposition of a Mo layer of $\sim$100 nm molybdenum. The film on the copper surface acts like a smooth layer improving the roughness of the surface. Actually, our tests confirm that the initial low roughness of the Cu machined surface is preserved or improved by Mo sputtering. However, since Cu and Mo have different thermal coefficients, the molybdenum deposited layer on top of copper is not stable vs. temperature. After deposition, to stabilize the Mo coating we performed a dedicated thermal treatment to fix the film on the substrate \cite{bini11}. The procedure allowed achieving homogeneous coatings and a reasonable contact force between the Mo layer on top of copper. 
To obtain information regarding the chemical composition of the coating as a function of the depth profile of the deposited molybdenum we used the Rutherford backscattering spectroscopy (RBS) \cite{bini11}. As an example, the depth profile analysis of a 350 nm molybdenum film is showed in figure \ref{fig:Fig3} as obtained after a thermal treatment at 600 C for 2 hours. The curve shows the atomic concentration of the relevant components of the deposited film vs. the distance from the sample surface. Data indicate that the film is uniformly contaminated vs. depth by oxygen ($\sim$ 20\%). %
Similar results have been obtained by photoemission spectroscopy data \cite{bini11}.

\section{XAS (X-ray Absorption Spectroscopy) characterization}

\begin{figure}[h]
\centerline{\includegraphics[scale=0.8]{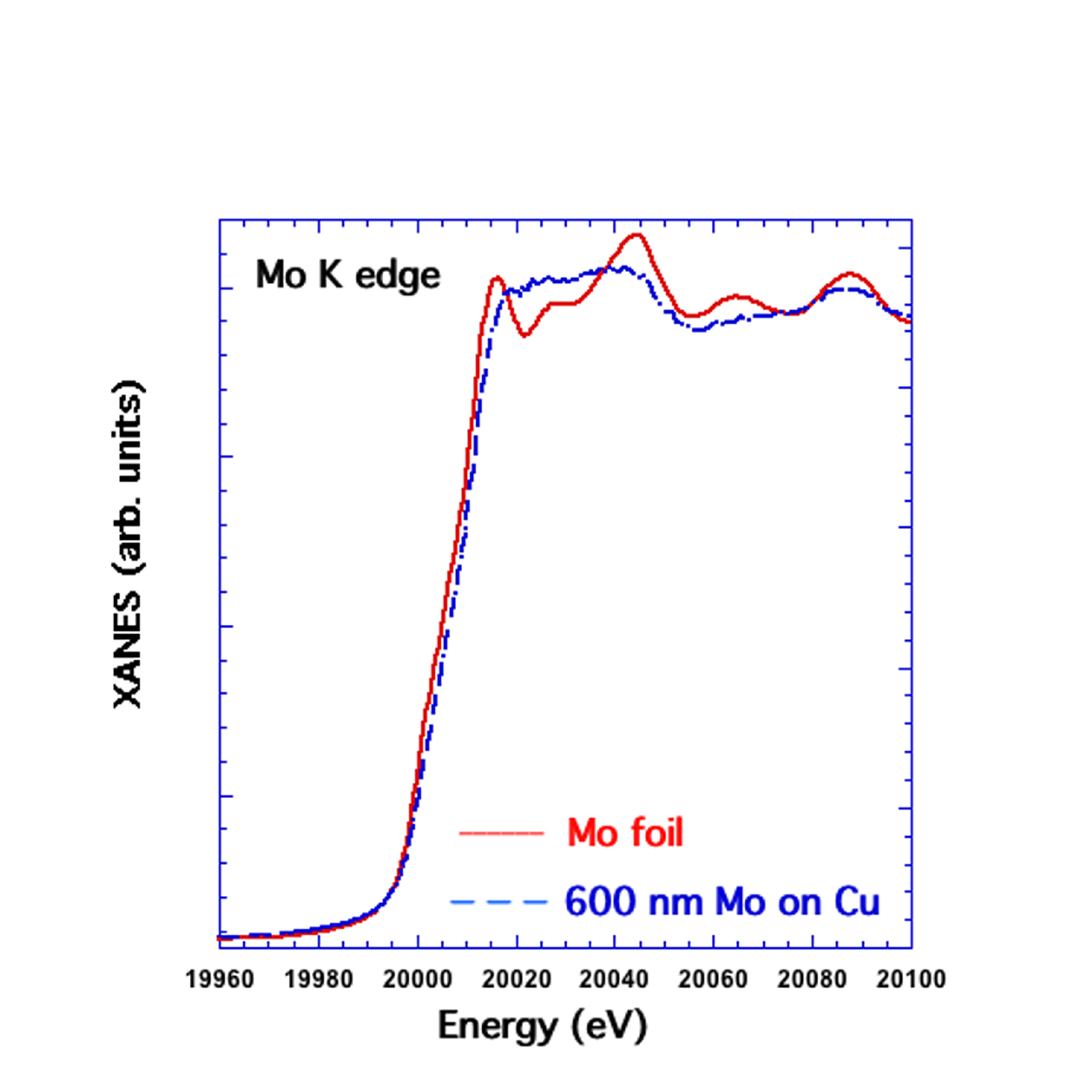}}
\caption{Comparison of spectra at the Mo K-edge between a Mo film (red) and a Mo coating on a copper disk (blu).}
\label{fig:Fig5}
\end{figure}

To characterize the chemical status of Mo atoms we performed XAS (X-ray Absorption Spectroscopy) experiments at the Mo K edge, moreover, working at grazing incidence we may enhance the signal associated to thin surface layers. X-ray Absorption Spectroscopy is a well-established technique that via methods such as EXAFS (Extended X-Ray Absorption Fine Structures) or XANES (X-ray Absorption Near Edge Structure) may return a direct quantitative measurement of local structure properties such as geometry, bond lengths and coordination numbers around a selected photo-absorber. With EXAFS we are also able to recognize the nature of the nearest neighbors around a photo-absorber atom while using XANES spectroscopy, a technique that looks at the intense features present in a short energy range around the atomic edge, we may obtain geometrical information of a small cluster around the photoabsorber.  It has been well demonstrated that using XAS spectroscopy a series of spectral characteristics, e.g., the position of the absorption threshold, depend by the electronic properties of the photoabsorber (in our case Mo) and therefore by its state of oxidation. The shape of the spectrum is characteristic of an atomic coordination, e.g., tetrahedral or octahedral, and also the presence of pre-edge structures may clarify coordination and oxidation of a photo-absorber. \\
Experiments at the Mo K-edge have been performed at Diamond on the beamline B18 using a Si(111) monochromator coupled to a Pt coated mirror focusing radiation in a spot of $\sim$200x200 $\mu$m \cite{Dent09}. Fluorescence radiation has been measured by a 9-elements Ge detector and a readout electronics XSPRESS2 collecting $\sim$300 kcs/sec per element. The incoming radiation has been monitored using an ionization chamber filled by Ar gas. 
To optimize signals from Mo atoms, experiments have been performed both at grazing incidence and at normal incidence in different areas of the film always in the fluorescence mode. As an example, in figure \ref{fig:Fig5} we show XANES data at the Mo K-edge of a Mo coated film ($\sim600$ nm) on a Cu disk of $\sim$2 cm of diameter. 
From these data, analyzing the shift of the energy position of the absorption threshold it is possible identify the oxidation state of Mo. Looking at figure \ref{fig:Fig5} the comparison of XANES spectra between a Mo film measured in the transmission geometry and a 600 nm Mo coated layer on a copper disk measured at grazing incidence, we may point out that the Mo coating has a slightly disordered structure (see the smoothed features at the edge) . Moreover, from the comparison with the available XANES spectra of different Mo oxides \cite{Ressler02} we may also recognize that these coatings are made by Mo dioxide (MoO$_2$) i.e., a Mo atom coordinated by four oxygen atoms. The result is confirmed by the observed shift of the absorption edge ($\sim$3 eV) in agreement with Mo dioxide XAS data available in the literature. \cite{Cramer76, Xu2012} 
  
\section{Electrical characterization}

\begin{figure}[h]
\centerline{\includegraphics[scale=0.8]{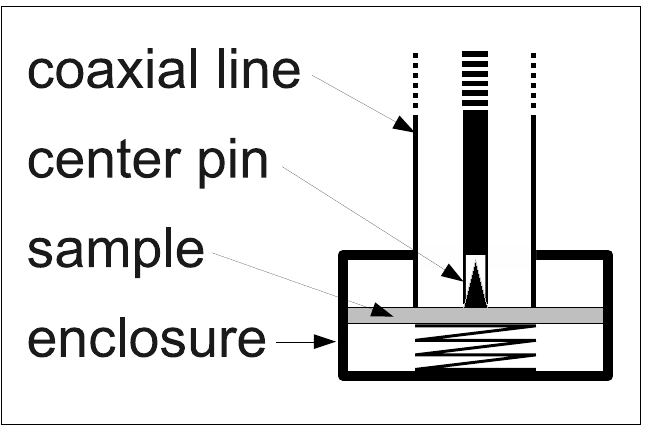}}
\caption{Sketch of the contact between sample and coaxial line.}
\label{fig:CorbinoLine}
\end{figure}

The electrical properties of Mo sputtered coatings has been performed via sample surface impedance probed as a function of the frequency through a Corbino disk geometry. In the latter the sample shortcircuits the coaxial cable connecting the sample to the instrument: a Vector Network Analyzer Anritsu 37297D. The connection between the cable and the sample was made through a double spring method, extensively described in \cite{toso1} (see figure \ref{fig:CorbinoLine}). The measured quantity is the microwave reflection coefficient, that is the complex ratio between the reflected wave and the incident one, measured at the instrument port. This coefficient can be related through standard line calibration to the reflection coefficient at the interface between the connector and the sample under measurement. Indeed, the calibration process is possible only down to the plane which separate the coaxial cable to the connector depicted in the figure. The very last section of the line, that is the connector, cannot be fully calibrated, and a different approach has to be used. The connector used is a commercial V103F (Anritzu) launcher, whose inner and outer diameters are 0.7 and 1.4 mm respectively. Comparing these numbers to the lateral dimensions of the samples, it is evident that only a rather small section of the sample is probed. For this reason, we may approximate the sample as an infinite layer extended in the directions perpendicular to the cable.

\subsection{Mode Matching (MM) Analysis}

It is possible to demonstrate that quite generally the reflection coefficient at the interface between the connector and the sample can be written as
\begin{equation}
\label{gamma_gen}
\Gamma=\frac{Y_0-Y_{eff}}{Y_0+Y_{eff}}
\end{equation}
where $Y_0$ is the line admittance and $Y_{eff}$ is an effective admittance of the sample, which depends on the resistivity as well as on the geometry of the sample. The expression of $Y_{eff}$ is in general quite complex (especially at high frequencies). For the special case of a sample infinitely extended in the directions perpendicular to the wave propagation (the so-called "open ended" geometry), the effective admittance has been calculated by Ganchev et al. \cite{ganchev}. Their result can be written in the following form: 
\begin{equation}
\label{Yeff_Ganchev}
Y_{eff}=\int_0^\infty K(\zeta,k_0)Y_s(\zeta)d\zeta
\end{equation}
where $K(\zeta,k_0)=k_0s_{ab}(k_0\zeta)$ with $s_{ab}(x)= [J_0(x b)-J_0(x a)]^2/(x\ln(b/a))$, $J_0$ is the Bessel function and $a$ e $b$ are the inner and outer radius of the coaxial cable,  while $Y(\zeta)=\omega\tilde\epsilon/k(\zeta)$, with $k(\zeta)=k_0\sqrt{\tilde\epsilon_r-\zeta^2}$. In these expressions $k_0=\omega/c$ (with $c$ the speed of light in vacuum and $\omega=2\pi\nu$) and $\tilde\epsilon$ is the (relative) complex permittivity of the material under study which, for a solid material at microwave frequencies is usually written as
\begin{equation}
\label{epsilon}
\tilde{\epsilon}=\epsilon_r-i\frac{1}{\epsilon_0\rho\omega}
\end{equation}
where $\epsilon_r$ is the relative dielectric constant and $\rho$ the dc resistivity\footnote{Due to the term $\epsilon_0=8.85\times10^{-12}$. Except very specific cases, it is evident that even in a moderately conducting material the second term largely dominates the first}. If the sample is a relatively thin film grown on a substrate, the expression for the effective admittance must take into account also further reflections that occurs at the "end" of the sample, that is at the interface between the film and the substrate. In the "open ended" geometry \cite{ganchev} the expression for the effective admittance changes to
\begin{equation}
\label{Yeff_layer_Ganchev}
Y_{eff}=\int_0^\infty K(\zeta,k_0)\left\{Y_s(\zeta)\frac{\bar{Y}_b(\zeta)+iY_s(\zeta)\tan[k_s(\zeta)d_s]}{Y_s(\zeta)+i\bar{Y}_b(\zeta)\tan[k_s(\zeta)d_s]}\right\}d\zeta
\end{equation}
where the subscript "s" refers to parameters of the sample and the subscript "b" refers to parameters of the "backplate", that is the substrate. $\bar{Y}_b(\zeta)$ is the effective admittance of the substrate, which has an expression analogous to the expression within parentheses in equation \ref{Yeff_layer_Ganchev} (in the special case of an infinitely thick substrate, $\bar{Y}_b(\zeta)=\omega\epsilon_b/k_b(\zeta)$).

\subsection{Calibration of the line}

\begin{figure}[h]
\centerline{\includegraphics[scale=0.45]{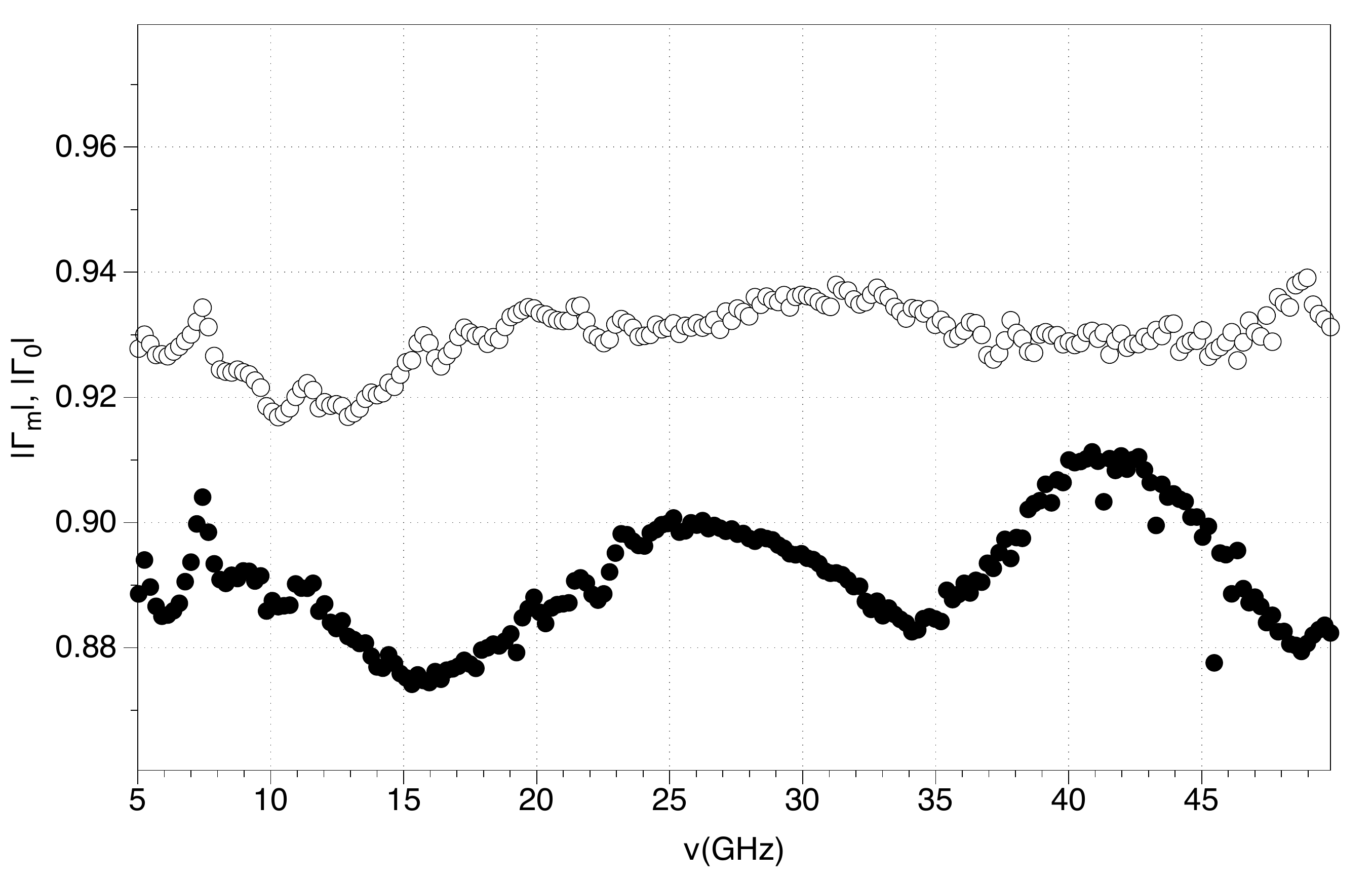}}
\caption{Measurements on the Mo film grown on glass substrate (see below). Filled symbols are raw data, open symbols the reflection coefficient at the film surface as obtained through the calibration procedure described in the text.}
\label{figmo-glass}
\end{figure}

Although a full calibration of the last section of the line (that is the connector) is not possible, { a partial calibration can be made by using three "known" samples. In fact, }we recall that quite generally the reflection coefficients measured on two different planes of a microwave line are related to each other by the relation
\[
\Gamma_1(\nu)=E_d(\nu)+\frac{E_r(\nu)\Gamma_2(\nu)}{1-E_s(\nu)\Gamma_2(\nu)}
\]
where $E_r$, $E_s$ and $E_d$ are frequency dependent, complex coefficients related to the properties of the segment of the line between the two sections 1 and 2. In particular, $E_r$ accounts for the attenuation and phase shift of the microwave during the travel across the line segment, $E_s$ and $E_d$ take into account possible reflections of the microwave signal within the line segment. Since none of the three coefficient is known {\em a priori}, a full calibration would require the {\em measurement} of three curves $\Gamma_1(\nu)$ and the {\em calculation} of three curves $\Gamma_2(\nu)$. Two of these curves can be obtained by measuring a polished copper plate (which is a good approximation of an infinitely conductive sample) and the glass side of one of the Mo/glass samples (whose calculated corresponding curve depends almost only on the glass parameters). {For the third curve, we use one of the Mo samples grown on a glass substrate, whose thickness and dc resistivity has been measured independently.}\\
 {This procedure would provide a full calibration of the line if it were possible to measure these samples with very high precision and to exactly know the parameters defining the electric response of the three measured samples. However, due to the non perfect contact between the connector and the measured sample, the measurement cannot be as precise as it would be requested by a calibration procedure. More, the parameters describing the samples are known only to a relatively low precision, introducing further source of possible error. As a result, this calibration cannot be as precise as the calibration of the remaining part of the line, where connections between line segments are realized through high precision connectors.}\\
The result of this procedure are the curves represented with open dots in figure \ref{figmo-glass}. Data point out that oscillations are due to the connector and after calibration reliable resistivity values can be obtained

\subsection{Results} 
 
In our coated films, although at priori we cannot exclude the presence of a thin oxide layer between the sample and the substrate we clearly have a Mo-based film grown on a copper substrate [9]. In this case a detailed measurement of the Mo resistivity is extremely difficult. In fact, it can be easily shown that, due to the thickness of the Mo film, $\tan[k_s(\zeta)d_s]\ll1$ so that, since further $Y_s<Y_b$, equation \ref{Yeff_layer_Ganchev} can be approximated by
\[
Y_{eff}\simeq\int_0^\infty K(\zeta,k_0)\frac{\bar{Y}_b(\zeta)}{1+i(\bar{Y}_b(\zeta)/Y_s(\zeta))\tan[k_s(\zeta)d_s]}d\zeta
\]
which, neglecting very special cases, is of the same order of magnitude of $Y_{Cu}=\int_0^\infty K(\zeta,k_0)\bar{Y}_b(\zeta)d\zeta$. Since the latter is much larger than the vacuum admittance $Y_0$, inserting this result in the equation \ref{gamma_gen} lead immediately to the conclusion that for such a sample one always has $\Gamma\simeq -1$ for any value of resistivity of the Mo film \footnote{This result, derived here for thin Mo films, can be shown to be general: indeed,  if the film is very thick ($k_sd_s\gg 1$ or, equivalently, $d_s \gg \delta$), equation \ref{Yeff_Ganchev} holds and, again, the effective admittance of the film is found to be much higher than $Y_0$}. 

\begin{figure}[h]
\centerline{\includegraphics[scale=0.48]{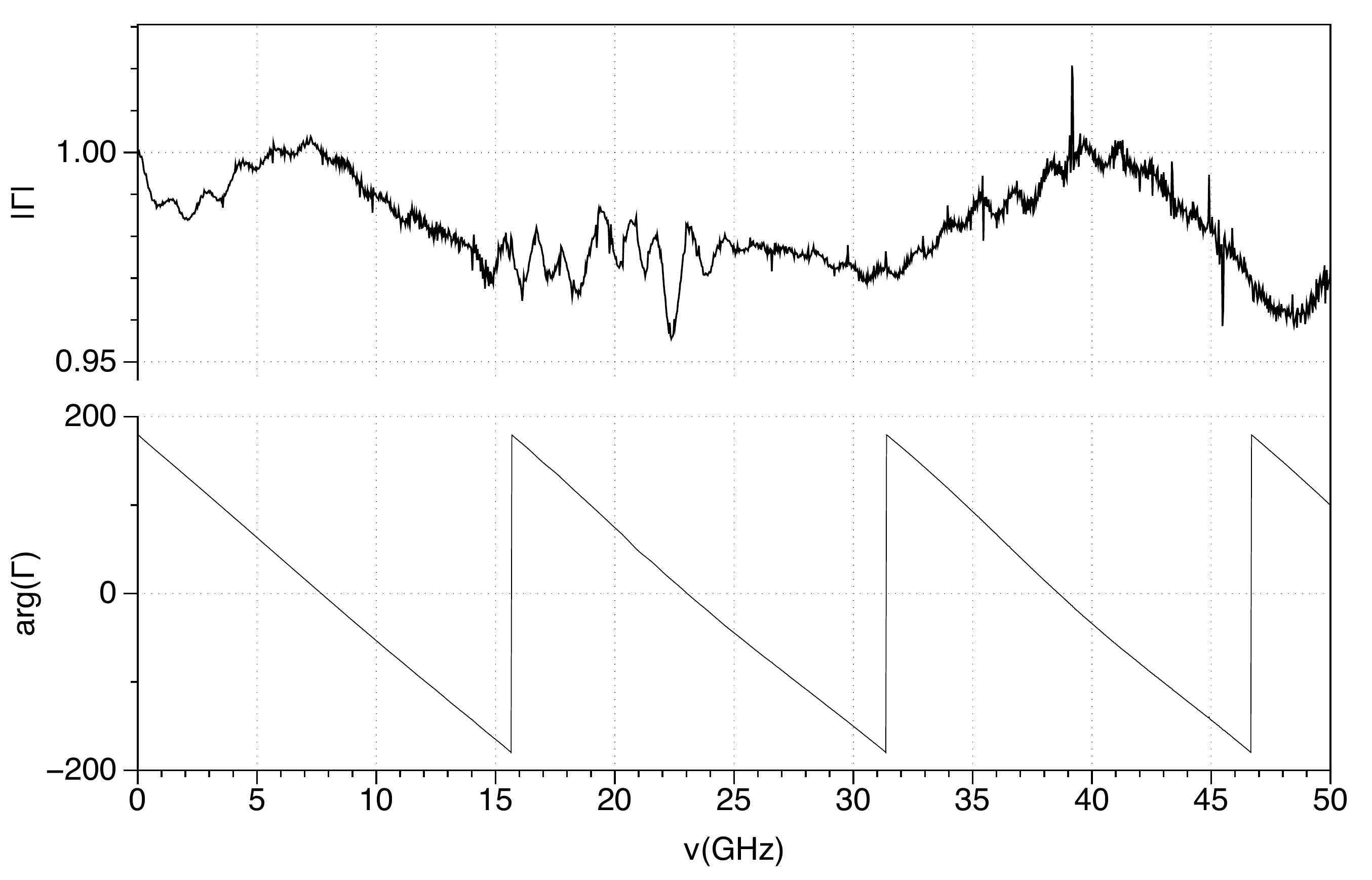}}
\caption{Modulus (upper panel) and phase (lower panel) of the uncalibrated reflection coefficient for a Mo thin film grown on Cu substrate.}
\label{Mo+Cu}
\end{figure}

This result is confirmed by measurements on thin Mo films grown on Cu substrate: as can be seen in figure \ref{Mo+Cu}, the modulus of the {uncalibrated} reflection coefficient is very close to one, while the behaviour of the phase as a function of frequency is due to the short uncalibrated section of the line.
In order to estimate the microwave resistivity of the Mo film, we have performed experiments on Mo films (3000, 6000 and 9000 $\AA$ thick) grown on Corning glass and (0001)-oriented single crystal Al$_2$O$_3$ substrates, starting from a 4 inches Mo target (99.99\%) in Ar (99.9999) atmosphere. The substrates were cleaned with proper solvents in ultrasonic bath, then they were dried with nitrogen flux. A 30 minutes pre-sputtering operation was performed before starting the Mo deposition. The substrate/target distance was 8 cm and the sputtering parameters were the following: background vacuum $4\times 10^{-7}$ Torr,  rf power 60 W, ambient temperature, pressure during the deposition 4 mTorr, Ar 23 sccm.  { A calibration curve of film thickness vs. sputtering time at fixed rf power, has been evaluated measuring the film thickness through the optical transmission spectrum.A deposition rate of 300 \AA/min was adopted}. The films were highly adhesive to the substrates as confirmed by a rudimentary tape test.

\begin{figure}[h]
\centerline{\includegraphics[scale=0.45]{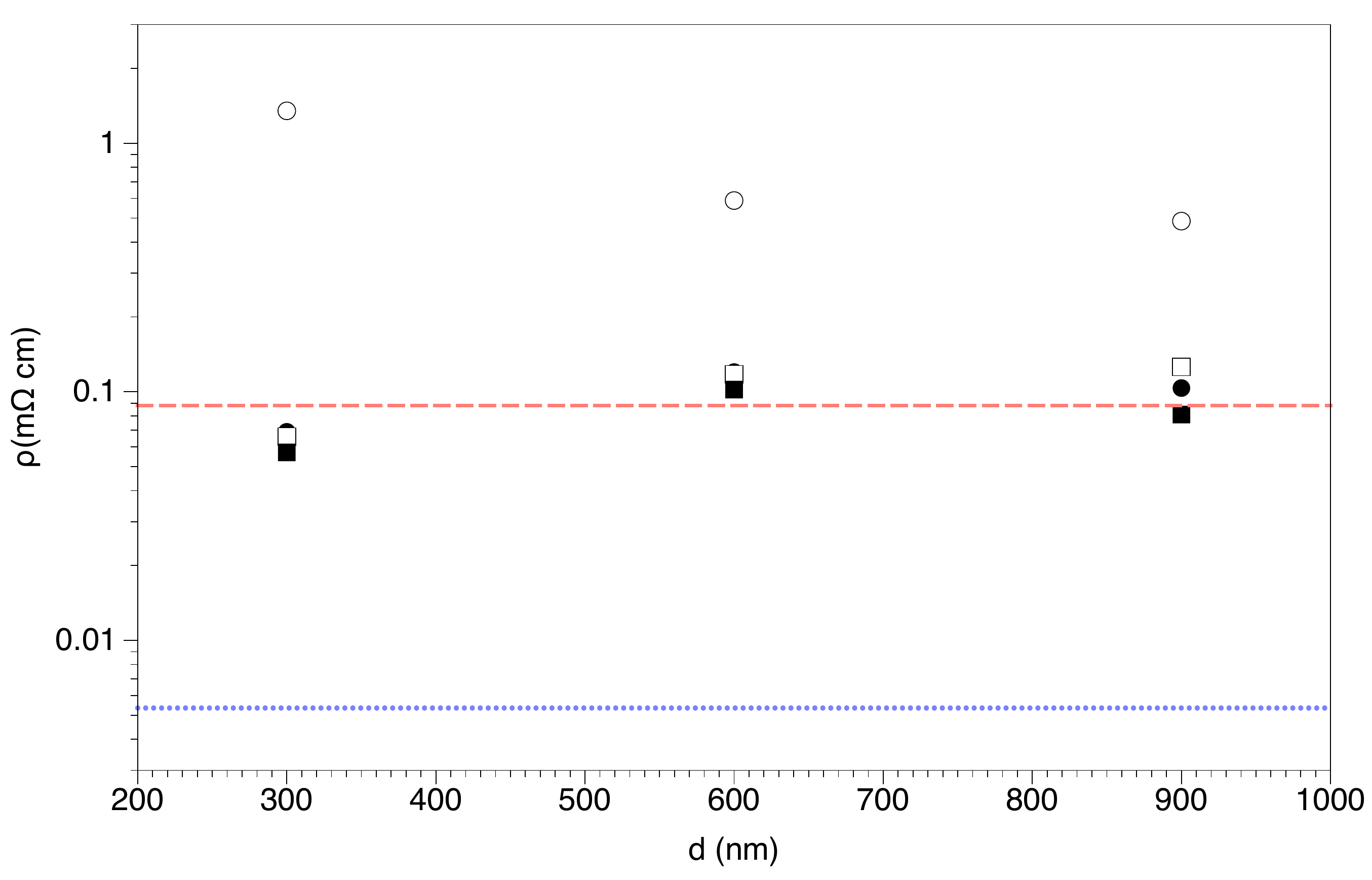}}
\caption{Values of resistivity measured on our samples. The resistivity of the first set of samples (open circles), where the presence of $MoO_2$ has been revealed though RBS and XANES, is sensibly higher than the resistivity of other sets. Also shown, for comparison, the literature values of resistivity for $MoO_2$ and pure $Mo$.}
\label{fig:resistivities}
\end{figure}

The values of resistivity obtained for the samples under study are reported in table \ref{rho_table} and in figure \ref{fig:resistivities}. Errors are estimated from the residual behavior as a function of frequency mainly due to the non perfect calibration of the line (figure \ref{figmo-glass}, open symbols).
{\em These values are compared to the literature values of the resistivity of pure Mo (0.005 m$\Omega$ cm) and of the Mo oxide (0.088 m$\Omega$ cm) in figure \ref{fig:resistivities}. Similar values available in literature \cite{Martinez:98}.}
\begin{table}[htdp]
\caption{resistivities of the measured samples}
\begin{center}
\begin{tabular}{|c|c|c|c|c|}
\hline
Substrate & $O_2$ pressure & thickness (nm) & $\rho$ (m$\Omega$ cm) & $\Delta\rho$ (m$\Omega$ cm)\\\hline
Glass & $10^{-2}$ & 300 & 1.4 & 0.1\\\hline
Glass & $10^{-2}$ & 600 & 0.59 & 0.05\\\hline
Glass & $10^{-2}$ & 900 & 0.49 & 0.05\\\hline
Glass & $1.5\times10^{-2}$ & 300 & 0.07 & 0.01\\\hline
Glass & $1.5\times10^{-2}$ & 600 & 0.12 & 0.01\\\hline
Glass & $1.5\times10^{-2}$ & 900 & 0.10 & 0.02\\\hline
Glass & $10^{-3}$ & 300 & 0.07 & 0.01\\\hline
Glass & $10^{-3}$ & 600 & 0.12 & 0.01\\\hline
Glass & $10^{-3}$ & 900 & 0.13 & 0.03\\\hline
$Al_2O_3$ & $10^{-3}$ & 300 & 0.057 & 0.003\\\hline
$Al_2O_3$ & $10^{-3}$ & 600 & 0.10 & 0.006\\\hline
$Al_2O_3$ & $10^{-3}$ & 900 & 0.08 & 0.02\\\hline
\end{tabular}
\end{center}
\label{rho_table}
\end{table}

It is clearly evident that the resistivity of samples belonging to the first series are much higher than those of the following series. 

{\bf Comparison with literature data of $Mo$ and $MoO_2$ points out a different contributions of molibdenum oxides that have to be characterized. Work is in progress to understand the differences and associate them to the coating procedures.There are many different types of molybdenum oxides with a different conductivity (see reference \onlinecite{Ressler02}). The resistivity values we measured are always larger than the literature values for corresponding bulk materials. This is certainly due to the different substrates, the amount of oxygen and the local structures as confirmed by XAS data.}\\

{\em Comparison with literature data of $Mo$ reveals however that in all cases the measured resistivity is much larger that the pure $Mo$. This is reasonably due to the fact that the structural characterization of the film reveal a relatively high degree of disorder and a non zero amount of oxygen.}\\

\begin{figure}[h]
\centerline{\includegraphics[scale=0.45]{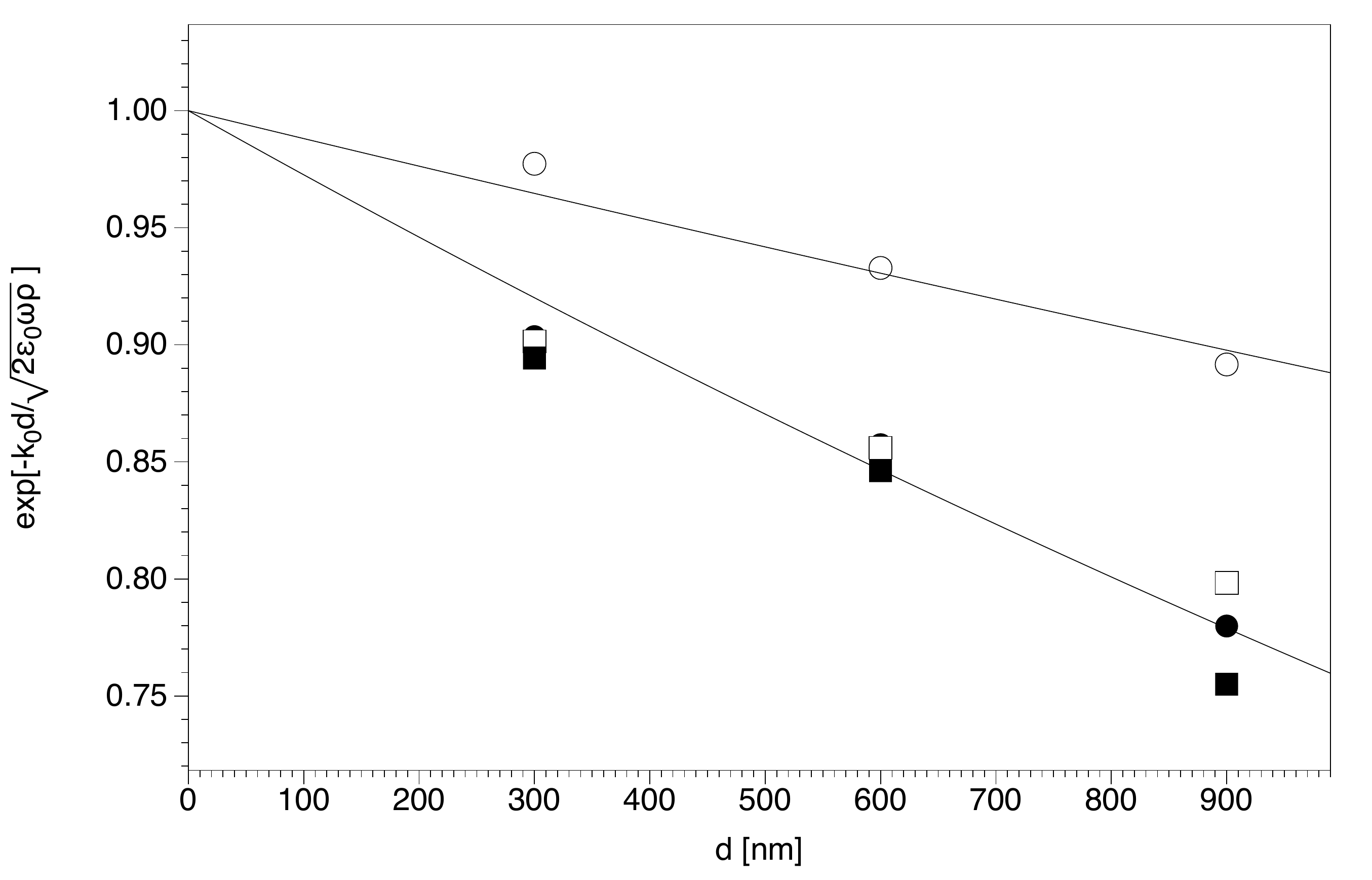}}
\caption{Ratio between the field at the substrate (glass) surface and the incident field at 10 GHz. Line represent the best exponential fit for the variation as a function of $d$, for samples belonging to the first series and for all other samples.}
\label{10GHz}
\end{figure}

With these values it is possible to estimate the ratio between the electric field at the glass surface and the incident field. This is done by calculating the attenuation through the Mo film, which is given by $\exp[-kd]$, with $k=k_0/\sqrt{(4\pi\epsilon_0\nu\rho)}$. This quantity is reported in figure \ref{10GHz} for all samples at $\nu=10$ GHz.

As can be easily seen in the figure, the field that reaches the substrate decreases as the thickness of the film increases. In addition, the lower resistivity observed for samples of series 2 to 4 results in a rather large decrease of the "residual" field. Extrapolating the observed attenuation to sample thicknesses of order 1$\mu$m, an attenuation as high as 25\% of the electric field (which correspond to more than 40\% of power) can be obtained before the substrate is reached.

\subsection{Conclusion} 

A large effort among INFN-LNF, KEK and SLAC laboratories to characterize and optimize Mo films deposited on Cu is in progress. We point out that molybdenum coatings obtained via sputtering is a promising approach to increase the accelerating gradient of accelerator cavities working at higher frequencies. In this contribution we describe chemical composition, deposition quality and resistivity properties of different molybdenum coatings on copper obtained via sputtering and stabilized by a dedicated thermal treatment. Chemical and electrical properties of these sputtered coatings have been determined by Rutherford backscattering, XANES and photoemission spectroscopies. However, a lot of work still has to be done to improve the quality and performances. Moreover, a full characterization and, in particular, conductivity properties and the behavior under high fields are still in progress because dedicated RF devices with such coatings have to be manufactured and tested at higher power. To optimize performances via coatings of these copper structures, deposition of other materials and different manufacturing methods and characterization of the Mo coatings are in progress. A SW structure made of copper cells, coated with Mo is ready is also under way.


\newpage

\begin{thebibliography}{99}

\bibitem{BiniCPC} S. Bini, V. Chimenti, A. Marcelli, L. Palumbo, B. Spataro, V. Dolgashev, S. Tantawi, Y. Higashi, M.G. Grimaldi, L. Romano, F. Ruffino, R. Parodi and A.D. Yeremian, Chinese Physics B (2012) in press

\bibitem{IPAC10}A. Gustafssono at al., Proceedings of IPAC'10, Kyoto, Japan

\bibitem{PRST09} A. Descoeudres, T. Ramsvik, S. Calatroni, M. Taborelli and W. Wuensh, DC breakdown conditioning and    breakdown rate of metals and metallic alloys under ultrahigh vacuum, PRST- Accelerators and Beams 12, 032001 (2009).

\bibitem{bini11}S. Bini et al, Development of X-Band Accelerating structures for high gradients, SPARC RF-11/004, 30/05/11

\bibitem{spataro11} B. Spataro et al., Technological issues and high gradient test results on X Band Molybdenum accelerating structures, NIMA (2011) doi: 10.1016/j.nima.2011.05.047

\bibitem{Higashi} Y. Higashi et al., Progress on Scanning Field Emission Microscope Development for Surface Observation, NIM A Proceedings D-11-00033R1

\bibitem{Martinez:98} M.A. Martinez, C. Guillen, Surface and coatings Technology 110 (1998) 62

\bibitem{Ressler02}	T. Ressler et al., J. Catal. 210, 67 (2002)

\bibitem{Dent09} A.J. Dent, G. Cibin, S. Ramos, A.D. Smith, S.M. Scott, L. Varandas, M.R. Pearson, N.A. Krumpa, C.P. Jones and P.E. Robbins, J. Physics: Conference Series 190 (2009) 012039

\bibitem{Cramer76} S.P. Cramer et al., J. Am. Chem. Soc. 98, 1287 (1976)

\bibitem{ganchev} Ganchev et al., IEEE Trans. Instr. Meas 44, 1023 (1995)

\bibitem{anlage} D.H.Wu, J.C.Booth, S.M.Anlage, \textit{Phys. Rev. Lett.}, {\bf 75}, 525 (1995)

\bibitem{toso1} N.Tosoratti, R.Fastampa, M.Giura, V. Lenzi, S.Sarti, E.Silva, \textit{Int. J. of Mod. Phys.} B, {\bf 14}, 2926 (2000). Note that using the notation of that paper $Z_0 = 1/Y_0$.

\end{thebibliography}
\end{document}